# Simultaneous Spin and Point-Group Adaptation in Exact Diagonalization of Spin Clusters

Shadan Ghassemi Tabrizi, Thomas D. Kühne

*Center for Advanced Systems Understanding (CASUS),*
*Am Untermarkt 20, 02826 Görlitz, Germany*
s. ghassemi-tabrizi@hzdr.de

**Abstract.** While either spin or point-group adaptation is straightforward when considered independently, the standard technique for factoring isotropic spin Hamiltonians by the total spin $S$ and the irreducible representation $\Gamma$ of the point-group is limited by the complexity of transformations between different coupling-schemes that are related by site-permutations. To overcome these challenges, we apply projection-operators directly to uncoupled basis-states, enabling the simultaneous treatment of spin and point-group symmetry without the need for recoupling-transformations. This provides a simple and efficient approach for the exact diagonalization of isotropic spin-models that we illustrate with applications to Heisenberg spin-rings and polyhedra, including systems that are computationally inaccessible with conventional coupling-techniques.

## 1. Introduction

With the emergence of the field of molecular magnetism [1,2], the rapid expansion in the synthesis of exchange-coupled clusters, along with their physical and spectroscopic characterization – such as temperature-dependent magnetic susceptibilities, electron paramagnetic resonance and inelastic neutron-scattering spectra – has made theoretical modeling based on effectively treating the open-shell metal ions as spin centers increasingly important [3,4]. The most common description is the Heisenberg model of pairwise scalar coupling, $\hat{H} = \sum_{i<j} J_{ij} \hat{\mathbf{s}}_i \cdot \hat{\mathbf{s}}_j$, where $\hat{\mathbf{s}}_i = (\hat{s}_{i,x}, \hat{s}_{i,y}, \hat{s}_{i,z})^T$ is a local-spin vector. The levels of such an isotropic Hamiltonian are characterized by a total-spin value $S$ and comprise $2S+1$ states with magnetic quantum numbers $M = -S, -S+1, ..., +S$ (eigenvalues of the $z$-component $\hat{S}_z = \sum_i \hat{s}_{i,z}$). The molecular symmetry manifests in terms of permutations of spin sites [5], allowing each level to be additionally assigned to an irreducible representation (irrep) $\Gamma$ of the respective point group (PG). Given the rapidly increasing dimension of the Hilbert-space with the number of centers, it becomes essential, when aiming for exact diagonalization (ED), to exploit all available symmetries to factor the Hamiltonian into smaller blocks, each characterized by labels $S$ and $\Gamma$ [6]. Beyond facilitating computations, a classification of levels can be useful for deriving selection rules [7,8] or for separating geometric and dynamic features of spectra [7–9].

Combining PG- and $S_z$-symmetry by working in a basis of uncoupled configurations $|m_1, m_2, ..., m_N\rangle$ defined by local $z$-projections with $M = \sum_i m_i$ is straightforward [5,10]. In addition, for $M = 0$, spin-flip symmetry can be used to separate sectors with even or odd $S$ [10]. Such approaches have been applied widely, e.g., in full ED of symmetric polyhedra [11] or in calculations of thermodynamic properties of lattice-fragments with up to $N = 42$ sites using the finite-temperature Lanczos method [12]. However, factorization of the Hamiltonian in terms of $\Gamma$ and $S$, which involves constructing eigenfunctions of $\hat{\mathbf{S}}^2$ (rather than only $\hat{S}_z$), is considered more demanding and was pursued in just a few cases. The most prominent strategy for this task [5,6,13–15] is based on genealogical coupling (GC) [3,16], where individual spins are coupled into subunits of increasing size until a multiplet with total spin $S$ is obtained. The calculation of the Hamiltonian elements in such a basis invokes irreducible tensor-operator techniques and proceeds by successive decoupling through Wigner-9$j$ symbols [3]. However, a complication arises from PG-adaption [5]: for most systems, there is no compatible coupling-scheme, i.e., PG-operations may produce states belonging to different schemes, related to the original basis by the respective site-permutations, and transforming back can become prohibitively demanding as it requires the computation of a large number of square-roots and Wigner-6$j$ symbols [6,13]. Resorting to the use of a subgroup that enables the construction of a compatible coupling-scheme [5,6] to avoid such transformations presents two drawbacks: it neither produces the smallest-possible Hamiltonian-matrix blocks, nor does it provide comprehensive labeling of eigenstates.

GC is not the only method for constructing spin-eigenfunctions [16], but Sahoo et al. [17] briefly reviewed why various other techniques are also difficult to apply in conjunction with arbitrary point-groups. To overcome this limitation, these authors suggested transformations between uncoupled and valence-bond (VB) states. In the VB-basis, each site $s_i$ is replaced by $2s_i$ auxiliary spin-1/2 objects, $N - 2S$ of which are singlet-paired according to Rumer-Pauling rules, supplemented by the restriction that no pairs are formed within the same center. As PG-symmetrization and setting up the Hamiltonian would require complicated transformations between different diagrams, VB-states are transformed into the uncoupled basis (using Clebsch-Gordan coefficients) where the PG-projectors and the Hamiltonian have simple representations. Thus, coupled and uncoupled bases are utilized for spin- and PG-adaptation, respectively. This approach is extendable to fermionic systems [18] and can be applied to all types of point-groups, but it has not been widely adopted. Since full ED was only demonstrated for a rather small system of a cubic arrangement of 14 $s=\frac{1}{2}$ sites (the dimension of the largest symmetry-subspace was 219) [17], it remains uncertain how well the procedure would scale to more challenging cases.

Our present strategy of applying spin- and PG-projectors to uncoupled states is suitable for arbitrary point-groups, and it is simpler than the GC- or VB-methods. Note that Bernu et al. [19] combined Löwdin's spin-projection operator [20] – $\hat{P}_S$ in Eq. (1) below – with a spatial-symmetry projector in a Lanczos-process to compute a small number of lowest states for clusters of up to 36 spin-1/2 centers. We recently used a similar approach to obtain analytical solutions for particularly small clusters [21]. In contrast, the present aim is the full numerical diagonalization of systems that would be challenging to solve without using all available symmetries. Details on forming spin-eigenfunctions based on a group-theoretical formulation of $\hat{P}_S$ (a procedure that we investigated recently [22]) and on the construction of the Hamiltonian in each $(\Gamma, S)$-subspace are provided in the following Theory and Computational Details section. In the Results and Discussion section, we assess the numerical accuracy of this approach and present several cases that represent some of the largest fully-solved spin-models to date, including systems that cannot be practically handled using the conventional GC-method.

## 2. Theory and Computational Details

In the following, we explain the combined spin- and PG-adaptation of uncoupled configurations to set up a generalized eigenvalue-problem (GEP) defined by the Hamiltonian and the overlap-matrix in a symmetry-subspace characterized quantum numbers $S$ and $\Gamma$. We start with a discussion of PG-adaptation. The projector onto a specific component $\mu$ of a possibly multidimensional irrep $\Gamma$ is given in Eq. (1),

$$\hat{P}_{\mu}^{\Gamma} = \frac{1}{h}\sum_{g}\left[D_{\mu\mu}^{\Gamma}(g)\right]^{*}\hat{G}(g), \qquad (1)$$

where $h$ is the order of the group, the sum runs over all elements $g$, $\hat{G}$ is a symmetry-operation that permutes the z-projections $|m_1,...,m_N\rangle$, and $D_{\mu\mu}^{\Gamma}$ is a diagonal element of the representation matrix $\mathbf{D}^{\Gamma}$ (the asterisk * denotes complex conjugation). Note that, when employing characters $\chi^{\Gamma} \equiv \mathrm{Tr}(\mathbf{D}^{\Gamma})$ instead of specific matrix-elements $D_{\mu\mu}^{\Gamma}$, components are not separated, see Eq. (2).

$$\hat{P}_{\Gamma} = \frac{1}{h}\sum_{g}\left[\chi^{\Gamma}(g)\right]^{*}\hat{G}(g) \qquad (2)$$

We work with Eq. (1) instead of Eq. (2) in order to reduce the size of the Hamiltonian-matrix in the respective sector by the number of components $n$ (the dimension of $\Gamma$), i.e., $\dim(\Gamma,\mu,M) = \frac{1}{n}\dim(\Gamma,M)$. The subspace dimension can be represented as the weighted sum of Eq. (3), where $\mathbf{P}_{\Gamma}$ is the matrix-representation of $\hat{P}_{\Gamma}$ in the basis with the selected $M$. To compute the traces of individual operations, $\mathrm{Tr}[\mathbf{R}(g)]$, we examine each $|m_1,...,m_N\rangle$ and check whether it is mapped onto itself under the action of $\hat{G}(g)$.

$$\dim(\Gamma,M) = \mathrm{Tr}(\mathbf{P}_{\Gamma}) = \frac{1}{h}\sum_{g}\chi^{\Gamma}(g)\mathrm{Tr}[\mathbf{R}(g)] \qquad (3)$$

The irreps of the cyclic group $C_N$, which is relevant for spin-rings, are one-dimensional and characterized by a crystal-momentum $k = 0, 1, ..., N-1$ associated with the eigenvalues $e^{-i2\pi k/N}$ of a site permutation $\hat{C}_N$ corresponding to a rotation by $2\pi/N$. Except for $k = 0$ (Mulliken label A) and $k = N/2$ (label B, for even $N$), $k$ and $N-k$ comprise degenerate pairs spanning two-dimensional irreps ($E_1$, $E_2$, etc.) of the dihedral group $D_N$. Therefore, when working with the $C_N$ group, ED can be restricted to the sectors $k = 0, 1, ..., \frac{N}{2}$. A few guidelines for simply constructing the required matrices for more complicated groups like icosahedral $I_h$ or octahedral $O_h$ were provided in our previous work [22]: permutation- and representation-matrices [23] are explicitly set up only for the group-generators, and the permutations are pairwise compounded until a closed set of elements is obtained; whenever a new element is found, the corresponding pair of representation-matrices is multiplied similarly to obtain the new representation-matrix.

In many cases – including in cyclic groups – a complete and orthogonal basis in a $(\Gamma, \mu, M)$ sector can be constructed by ensuring that each uncoupled state appears in at most one linear combination. The configurations $|m_1,...,m_N\rangle$ selected by this rule form a complete orthogonal space $(\Gamma, \mu, M)$ upon application of $\hat{P}_\mu^\Gamma$ and are stored in the columns of the matrix $\tilde{\mathbf{R}}$ (each column of $\tilde{\mathbf{R}}$ has exactly one entry of 1). While it was previously assumed that each $|m_1,...,m_N\rangle$ would always contribute to at most one linear combination [21], we have observed that this does not generally hold for multidimensional irreps of the icosahedral group, where a $|m_1,...,m_N\rangle$ may appear in multiple states for a given $(\Gamma, \mu, M)$, provided its coefficients have different magnitudes in different states. A small threshold on the number of these distinct absolute amplitudes (e.g., two or three) suffices to generate the complete $(\Gamma, \mu, M)$ space. The resulting $\tilde{\mathbf{R}}$ is generally overcomplete with respect to $\hat{P}_\mu^\Gamma$. However, rather than directly pruning linear dependencies, a single rank-revealing selection (as discussed below) is performed later to choose configurations for simultaneous spin- and PG-projection.

For large $\dim(\Gamma, \mu, M)$, we typically stop the generation of $\tilde{\mathbf{R}}$ once approximately 20% more states have been collected than the number $\dim(\Gamma, S)$ of levels. If the subsequent spin-adaptation step fails to fully span the $(\Gamma, S)$-space with this truncated set, then additional configurations $|m_1,...,m_N\rangle$ would be considered. However, we found that to produce a suitable truncated set, the $|m_1,...,m_N\rangle$ should be processed in random order, not sequentially.[1]

Turning now to the construction of spin-eigenfunctions, we note that direct application of Löwdin's projector, Eq. (4),

$$\hat{P}_S = \prod_{l \neq S} \frac{\hat{\mathbf{S}}^2 - l(l+1)}{S(S+1) - l(l+1)}, \quad (4)$$

can induce significant numerical rounding-errors, due to the repeated matrix-vector multiplications needed to isolate the subspace with the desired $S$ [19,22]. To circumvent these issues, we employ a group-theoretical integral-formulation of $\hat{P}_S$ [24],

$$\hat{P}_S = \frac{2S+1}{8\pi^2} \int_0^{2\pi} d\alpha \int_0^\pi d\beta \sin\beta \int_0^{2\pi} d\gamma \left[ D_{MM}^S(\alpha, \beta, \gamma) \right]^* e^{-i\alpha \hat{S}_z} e^{-i\beta \hat{S}_y} e^{-i\gamma \hat{S}_z}, \quad (5)$$

---

[1] For example, for $s = \frac{1}{2}$, a listing according to a binary representation, where $m_i = +\frac{1}{2}$ and $m_i = -\frac{1}{2}$ represent 1 and 0, would not be adequate; the ordering should be randomized for generating a truncated PG-adapted basis.

where $D_{MM}^{S}(\alpha,\beta,\gamma)$ is a diagonal element of the Wigner rotation-matrix for spin $S$. Evaluating the wave-function $\psi_{\{m_i'\},\{m_i\}}^{S}$, Eq. (6), obtained by applying $\hat{P}_S$ to a state $|m_1,...,m_N\rangle$ with $\sum_i m_i = M$, is greatly simplified by the integral formulation.

$$\psi_{\{m_i'\},\{m_i\}}^{S} \equiv \langle m_1',...,m_N' | \hat{P}_S | m_1,...,m_N \rangle \qquad (6)$$

The integrals over Euler angles $\alpha$ and $\gamma$ (corresponding to rotations about the $z$-axis) are trivial. The remaining integration over $\beta$ involves the element $d_{MM}^{S}(\beta)$ of the small Wigner-matrix and reduces to a standard-integral for spin-1/2 systems [25], which leads to a closed-form expression for $\psi_{\{m_i'\},\{m_i\}}^{S}$, Eq. (7), in terms of a special kind of so-called Sanibel-coefficients, involving only a sign-factor and a binomial coefficient,

$$\psi_{\{m_i'\},\{m_i\}}^{S} \propto (-1)^k \binom{N_\uparrow}{k}^{-1}, \qquad (7)$$

where $k$ is the number of sites with $m_i' < m_i$, and $N_\uparrow$ is the number of $\uparrow$ sites in the reference configuration, i.e., $N_\uparrow + N_\downarrow = N$, $M = \tfrac{1}{2}(N_\uparrow - N_\downarrow)$. For $s > \tfrac{1}{2}$, the wave function becomes a linear combination of standard-integrals [22].[2] As we recently demonstrated [22], this analytical group-theoretical approach to evaluate the action of $\hat{P}_S$ provides numerical advantages over Löwdin's projector. Accordingly, we exclusively employ this method in the present work.

The number of multiplets with spin $S$ transforming according to $\Gamma$ is calculated from Eq. (8), assuming $M \geq 0$. The $S_{\max} = \sum_i s_i$ level is symmetric under all spin-permutations; thus, $\dim(\Gamma_1, S_{\max}) = 1$, where $\Gamma_1$ is the totally-symmetric irrep.

$$\dim(\Gamma,S) = \dim(\Gamma,\mu,M) - \dim(\Gamma,\mu,M+1) \qquad (8)$$

A subset (see next section on the selection-procedure) of $\dim(\Gamma,S)$ columns from $\tilde{\mathbf{R}}$ comprises the matrix $\mathbf{R}$. These configurations span a complete, linearly independent $(\Gamma,S)$-basis under the application of the projectors. With $\mathbf{P}_S$ and $\mathbf{P}_\Gamma$ denoting the respective matrices,[3] and considering the Hermiticity and idempotency of projection-operators, the Hamiltonian and overlap-matrices in a $(\Gamma, S)$ sector are formulated in Eqs. (9) and (10).

---

[2]One may alternatively evaluate the integrals numerically on a grid.

[3]Here and in the following, to avoid clutter, we will typically just refer to $\Gamma$ for simplicity and let $\hat{P}_\Gamma$ denote $\hat{P}_\mu^\Gamma$, while keeping in mind that the projection is always performed onto a specific component $\mu$.

$$\mathbf{H}_{\Gamma,S} = \mathbf{R}^\dagger \mathbf{P}_S^\dagger \mathbf{H} \mathbf{P}_\Gamma \mathbf{R} \quad (9)$$

$$\mathbf{S}_{\Gamma,S} = \mathbf{R}^\dagger \mathbf{P}_\Gamma \mathbf{P}_S \mathbf{R} \quad (10)$$

The Hamiltonian $\mathbf{H}$ (in the uncoupled basis with $M = S$) is generated in sparse format and is multiplied by the PG-adapted states $\mathbf{P}_\Gamma \mathbf{R}$, which are also in sparse format. Spin-adaptation works as described above, cf. Eq. (7) for $s = \frac{1}{2}$, but as spin-eigenfunctions are dense vectors, fully storing $\mathbf{P}_S \mathbf{R}$ may require excessive memory. Therefore, a single row (or a limited number of rows) of $\mathbf{R}^\dagger \mathbf{P}_S^\dagger$ is multiplied with the entire $\mathbf{H} \mathbf{P}_\Gamma \mathbf{R}$ to construct $\mathbf{H}_{\Gamma,S}$ (as well as $\mathbf{S}_{\Gamma,S}$) sequentially. A symmetrization, $\mathbf{H}_{\Gamma,S} = \frac{1}{2}(\mathbf{H}_{\Gamma,S} + \mathbf{H}_{\Gamma,S}^T)$ and $\mathbf{S}_{\Gamma,S} = \frac{1}{2}(\mathbf{S}_{\Gamma,S} + \mathbf{S}_{\Gamma,S}^T)$ should be performed to eliminate numerical rounding-errors, which would slow down the subsequent solution of the GEP and could lead to small imaginary components of the eigenvalues. The nonorthogonality of the symmetry-adapted basis is a feature shared with the GC- and VB-methods mentioned in the Introduction [13,17]. Rather than performing an explicit orthogonalization, we solve the GEP with the `eig` function in `Matlab`.

We now discuss the selection of $\tilde{\mathbf{R}}$ states $\mathbf{P}_\Gamma$ from $\dim(\Gamma, S)$ for the combined application of $\mathbf{R}$ and $\mathbf{P}_\Gamma$. As explained, the columns of $\mathbf{P}_S$ span the (possibly truncated) space $(\Gamma, \mu, M)$ under $\mathbf{P}_\Gamma$. Note that for pure spin-projection in spin-1/2 systems, the so-called Löwdin-theorem specifies how configurations ought to be selected so that, under the action of $\mathbf{P}_S$, they span a complete, linearly independent basis in the $S = M$ sector [26]. We have recently generalized Löwdin's theorem to systems with arbitrary local spin-quantum numbers [22] but observed that it does not constitute the most numerically stable procedure in terms of the conditioning of the overlap-matrix (i.e., eigenvalues can approach zero within numerical precision). Instead, we proposed an iterative pivoted Cholesky-decomposition [27,28] (PCD). This strategy efficiently selects a linearly-independent subset for constructing a well-conditioned GEP. A practical alternative is a rank-revealing QR-factorization with column-pivoting, which is directly available in `MATLAB` without additional coding, contrary to PCD.

To our knowledge, there is no analogue of Löwdin's theorem for selecting configurations for the combined action of $\mathbf{P}_\Gamma$ and $\mathbf{P}_S$, and we thus build on the iterative strategy described previously to construct a $(\Gamma, S)$-basis [22]. However, this procedure is now applied in a slightly modified fashion:[4] rather than initially selecting a minimal number $\dim(\Gamma, S)$ of candidate states from $\tilde{\mathbf{R}}$, we choose roughly 10–20 % more to reduce the likelihood of multiple iterations.[5] From these selected states, provisional overlap- and Hamiltonian-matrices $\mathbf{S}_{\Gamma,S}$ and $\mathbf{H}_{\Gamma,S}$ are formed, and the rank $r$ of $\mathbf{S}_{\Gamma,S}$ is computed at a reasonable tolerance well-above the numerical accuracy. At each step, the PCD routine identifies the pivot – the index of the largest diagonal element of the current residual-matrix (initially set to $\mathbf{S}_{\Gamma,S}$) – then forms a normalized scaling-vector to update that matrix. This procedure is repeated until $r$ pivots have been recorded, as detailed in Figure 1. States failing the selection are removed, and if the rank of $\mathbf{S}_{\Gamma,S}$ in the subset of the selected pivot-indices remains below $\dim(\Gamma, S)$, then an additional batch of states is appended. If the process stagnates, the tolerance for rank-determination may be reduced. Each iteration only updates $\mathbf{S}_{\Gamma,S}$ and $\mathbf{H}_{\Gamma,S}$ for any newly added states, thereby avoiding a recomputation of matrix-elements.

```
pivots = [] % initialize empty list
T = S % initialize residual-matrix T in terms of overlap--matrix S
for k = 1 to r do
    [d,p] = max(diag(T))   % value (d) and index (p) of
                           % largest diagonal element
    pivots = [pivots,p]    % append current pivot to list
    v = T(:,p)/sqrt(d) % compute scaling-vector
    T = T – v*v† % update T; superscript † denotes conjugate
end
Return pivots
```

Figure 1: `Matlab`-like pseudocode for a pivoted Cholesky-decomposition (PCD). Starting from the overlap-matrix $\mathbf{S}$, the algorithm iteratively selects $r$ basis states to span a linearly-independent space by updating the residual-matrix $\mathbf{T}$ and tracking the pivot-indices.

In our implementation, the construction of spin-eigenstates by analytically evaluating the group-theoretical integral of Eq. (5) tends to represent the computational bottleneck for $s > \frac{1}{2}$.

---

[4] The explanation focuses specifically on PCD, but the iterative approach for state-selection could be performed similarly for QR-factorization with column-pivoting.

[5] If $\tilde{\mathbf{R}}$ is a complete set of PG-adapted states generated from an ordered list of uncoupled configurations, its columns should be selected randomly.

We therefore accelerate the construction of $\mathbf{S}_{\Gamma,S}$ and $\mathbf{H}_{\Gamma,S}$ by processing uncoupled states in parallel. With respect to assembling and solving the GEP, different $(\Gamma,S)$-sectors are independent – aside from the minor overhead of constructing the same uncoupled basis and Hamiltonian $\mathbf{H}$ for a given $S$ but different $\Gamma$ – thus permitting straightforward distribution of these computations across nodes of a compute-cluster.

## 3. Results and Discussion

We consider spin rings and polyhedra – specifically, the icosahedron, cuboctahedron and truncated tetrahedron – to illustrate the exact-diagonalization procedure using a complete symmetry-factorization of the Heisenberg spin-Hamiltonian. Throughout, we assume nearest-neighbor couplings with $J=1$, representing antiferromagnetic interactions.

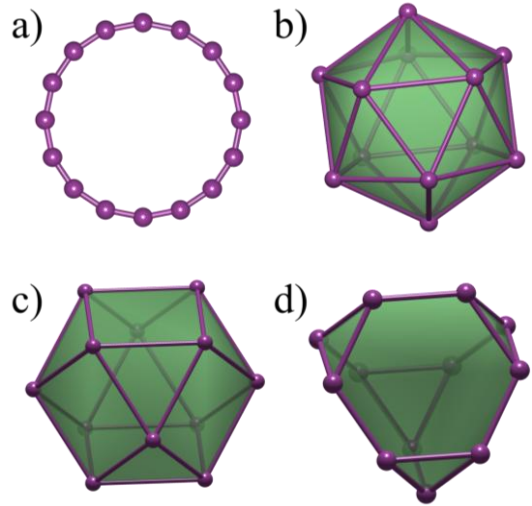

Figure 2: Heisenberg spin-clusters considered in this work: a) ring, b) icosahedron, c) cuboctahedron, d) truncated tetrahedron. Spheres denote spin-sites, and lines represent antiferromagnetic couplings with $J=1$.

**Spin-ring.** As a first example, we compare the results of two methods for factoring the spin-Hamiltonian for an $N=16$, $s=\tfrac{1}{2}$ ring: (1) using the straightforward PG-adaptation of uncoupled states, labeling sectors by $(\Gamma, M)$;[6] and (2) combining PG-adaptation with total-spin projection, labeling sectors by $(\Gamma, S)$. The first method involves a standard eigenvalue-problem, while the second requires solving a generalized eigenvalue-problem (GEP). The numerical accuracy of the $(\Gamma, M)$-approach is guaranteed to be at approximately machine-precision because all Hamiltonian-elements are computed using exact algebraic expressions in `Matlab`'s standard double-precision arithmetic. This ensures that the eigenvalues serve as a reliable benchmark for comparison with the $(\Gamma, S)$-based results. In the $(\Gamma, S)$-approach, the spin-projection is performed exactly using Sanibel-coefficients, as detailed above. This procedure avoids numerical problems that could arise when using Löwdin's projector or if the group-theoretical projection-integral were discretized over a grid. However, $(\Gamma, S)$-adaptation inherently involves solving a GEP, which may introduce errors due to the conditioning of the overlap-matrix and the limitations of the numerical solver. As noted in Theory and Computational Details, we verify that the rank of the overlap-matrix corresponds to the full dimension of the symmetry-subspace within a reasonable tolerance. Figure 3 demonstrates that the energy errors for all levels are negligibly small for any practical purpose, confirming the reliability of our $(\Gamma, S)$-adaptation.[7]

---

[6]Quantum numbers $S$ can be assigned to the levels using one of two methods: i) calculating spectra in all $M \geq 0$ subspaces; if a level appears in a specific $M$-subspace but not in $M+1$, it corresponds to $S = M$, or ii) diagonalizing only in the space with the smallest absolute value of $M$ ($M=0$ or $M=\tfrac{1}{2}$) and assigning $S$ from the expectation value $\langle \hat{\mathbf{S}}^2 \rangle = S(S+1)$, which requires access to the eigenfunctions.

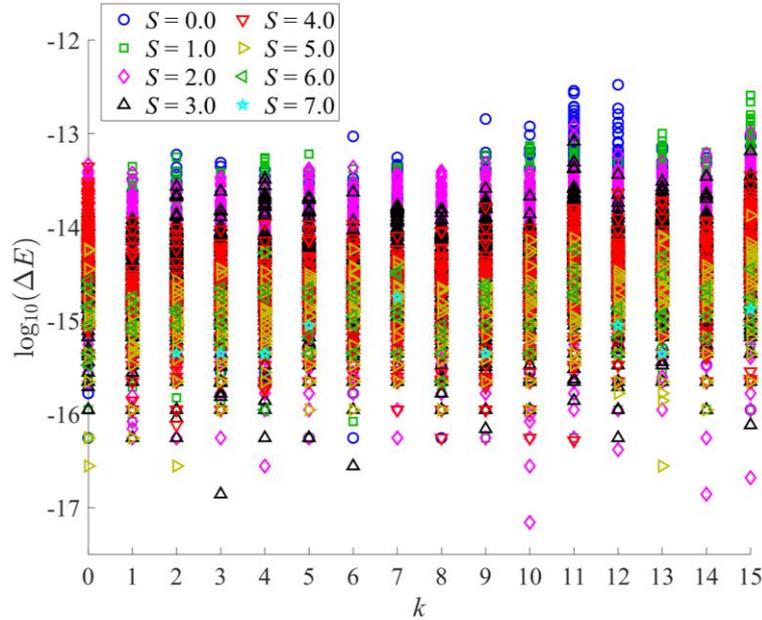

Figure 3: Comparison of energy differences from ED based on adaptation to $(\Gamma, M)$ or $(\Gamma, S)$, as a function of the crystal momentum $\Gamma = k$, for an $N = 16$, $s = \frac{1}{2}$ ring. The energy of the $S_{max} = 8$ level is exact in both calculations $(E = 4)$; therefore, $\log_{10}(\Delta E)$ is undefined.

---

[7] The numerical errors for degenerate irrep-pairs, such as $k = 1$ and $k = 15$, differ because the random selection of spin-configurations for applying the projectors is performed independently for each sector.

To illustrate the utility of the spectrum, we follow Ref. [13] by focusing on two thermodynamic properties: magnetization and magnetic heat-capacity as a function of temperature and magnetic field. Both quantities can be readily computed using the complete set of eigenvalues. The thermal average $\langle \mathcal{M} \rangle$ of the magnetization is given by Eq. (11):

$$\langle \mathcal{M} \rangle = \frac{1}{q} \sum_i g\mu_B M_i e^{-\beta E_i} \qquad (11)$$

where $M_i$ is the magnetic quantum number for state $i$, $q = \sum_i e^{-\beta E_i}$ is the molecular partition-function, $\beta = \frac{1}{kT}$, and $E_i = E_i^{(0)} - g\mu_B B M_i$ represents the total energy, which includes the field-free energy $E_i^{(0)}$ obtained from ED and a Zeeman-contribution that depends on the external magnetic field $B$, where $g$ and $\mu_B$ denote the isotropic $g$-factor and the Bohr-magneton, respectively.

The heat capacity $\mathcal{C}$, defined in Eq. (12),

$$\mathcal{C} = \frac{\partial \langle \mathcal{E} \rangle}{\partial T} , \qquad (12)$$

is expressed in terms of the internal energy $\langle \mathcal{E} \rangle = \frac{1}{q} \sum_i E_i e^{-\beta E_i}$ and can be calculated from the thermal energy-fluctuation, as shown in Eq. (13),

$$\mathcal{C} = \frac{\beta^2}{k} \left( \langle \mathcal{E}^2 \rangle - \langle \mathcal{E} \rangle^2 \right) , \qquad (13)$$

where $\langle \mathcal{E}^2 \rangle = \frac{1}{q} \sum_i E_i^2 e^{-\beta E_i}$. For the $N = 16$, $s = \frac{1}{2}$ ring, $\langle \mathcal{M} \rangle$ and $\mathcal{C}$ are plotted in Figure 4 and Figure 5, respectively, where we set $\mu_B = k = 1$ and $g = 2$.

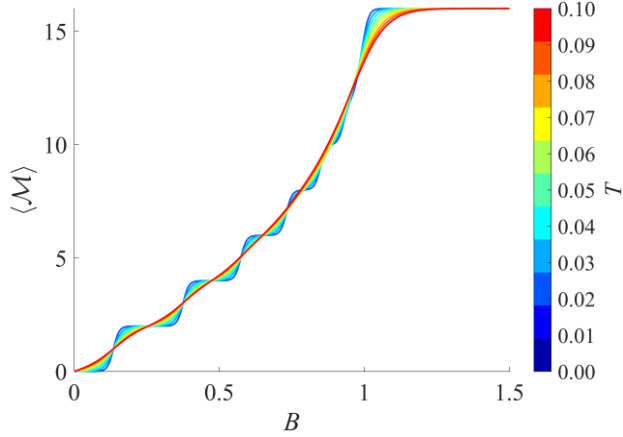

Figure 4: Magnetization $\langle\mathcal{M}\rangle$ as a function of the external field $B$ for various temperatures $T$ (indicated by the color bar) for an $N=16$, $s=\frac{1}{2}$ ring. Refer to the main text for further details.

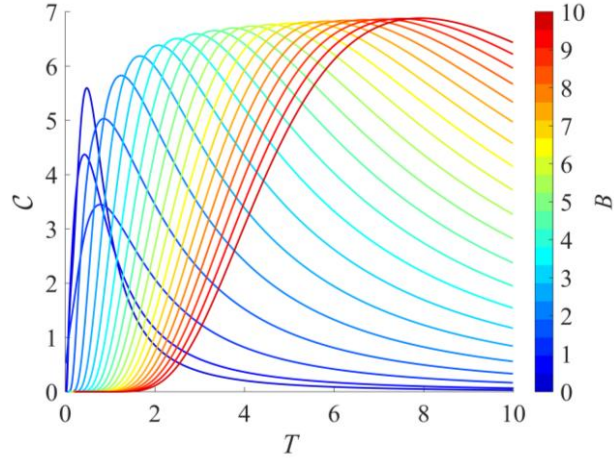

Figure 5: Magnetic heat-capacity $\mathcal{C}$ for an antiferromagnetic $N=16$, $s=\frac{1}{2}$ ring as a function of $T$ for various field-strengths $B$ (indicated by the color bar).

**Icosahedron.** The $s=1$ icosahedron has already been fully diagonalized by Konstantinidis, employing the $I_h$ point-group along with $S_z$ and spin-flip symmetries [11]. However, for an $s=\frac{3}{2}$ system, such an approach becomes challenging due to the significantly larger state-space dimension (cf. Table 2 in the Appendix). Although the Hamiltonian- and overlap-matrices in the $(\Gamma, M)$-basis are stored in a memory-efficient sparse format, exploiting this advantage for a complete diagonalization would require specialized libraries. The built-in `eig` function in MATLAB, which computes the full spectrum for (generalized) eigenvalue-problems, does not accept sparse inputs; sparse matrices must thus be converted into their full counterparts. On the other hand, the iterative solver `eigs` is designed to compute only a subset of eigenvalues (and eigenvectors) of sparse matrices.

Using the GC-method required three days on 128 processors to construct the Hamiltonian in the largest subspace in the $s=1$ icosahedron (dimension 3315, $S=2$, $\Gamma=H_g$), primarily due to the computationally demanding transformations between coupling schemes [6]. Using our current projection-method on a desktop computer with six CPU cores, setting up the GEP in the respective space[8] takes only a few seconds, and determining the full spectrum takes two minutes.[9] In our current implementation, the main bottleneck is the construction of spin-eigenstates, which requires evaluating the analytical form of the group-theoretical integral of Eq. (5). The spin-adaptation for $s > \frac{1}{2}$ is more demanding for $s = \frac{1}{2}$ systems of similar Hilbert-space dimension, because the integral – as explained in Ref. [22] – decomposes into a sum of analytically-evaluable standard-integrals, and the number of terms in the sum grows steeply as $s$ increases. In contrast, for $s = \frac{1}{2}$ systems, only a single standard-integral needs to be evaluated.

Even with 256 processors, a full adaptation with respect to total-spin and $I_h$-symmetry for the $s = \frac{3}{2}$ icosahedron could not be completed using the GC-approach [6]. Instead, this system was diagonalized by adopting the $D_2$-subgroup of $I_h$, which permits the construction of a compatible coupling-scheme; however, the Hamiltonian-blocks are significantly larger compared to full $I_h$-symmetry. With our approach, we are now able to adapt to $(\Gamma, S)$ within the complete $I_h$ point-group without any difficulty: the determination of the full spectrum of the $s = \frac{3}{2}$ icosahedron took us approximately 16 hours on a cluster node with 64 GB RAM and 8 CPU cores. The dimensions of the respective subspaces are collected in Table 4 in the Appendix and spectra are illustrated in Figure 6.

---

[8]As explained in the Theory and Computational Details section, we separate the five components of $H_g$, reducing the matrix dimension in the $S=2$, $\Gamma=H_g$ space to $3315/5 = 663$, see

Table 1 in the Appendix.

[9]This estimate is provided primarily for illustrative purposes, as it is likely possible to improve the implementation to reduce the runtime.

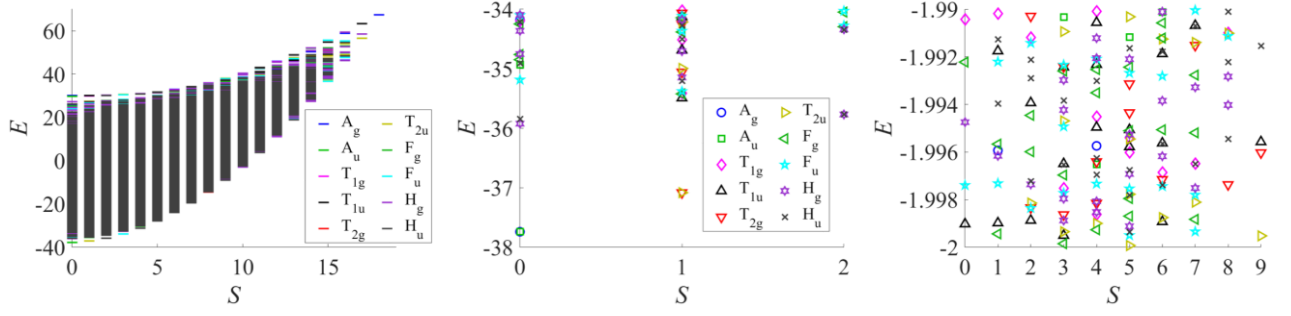

Figure 6: Full spectrum of the $s = \frac{3}{2}$ icosahedron (left), low-energy section (middle), and an arbitrary section from the intermediate energy-range (right).

The magnetic heat-capacities in zero-field for the icosahedron with $s = \frac{1}{2}$, $s = 1$ and $s = \frac{3}{2}$ are shown in Figure 7. Note that there are two peaks in the low-temperature part of the curve for the $s = \frac{3}{2}$ system. The first peak at $T \approx 0.008$ is due to a near-degeneracy of the ground-state, as the lowest levels in sectors $(S = 0, A_g)$ and $(S = 0, A_u)$ have $E = -37.739$ and $E = -37.741$, respectively.

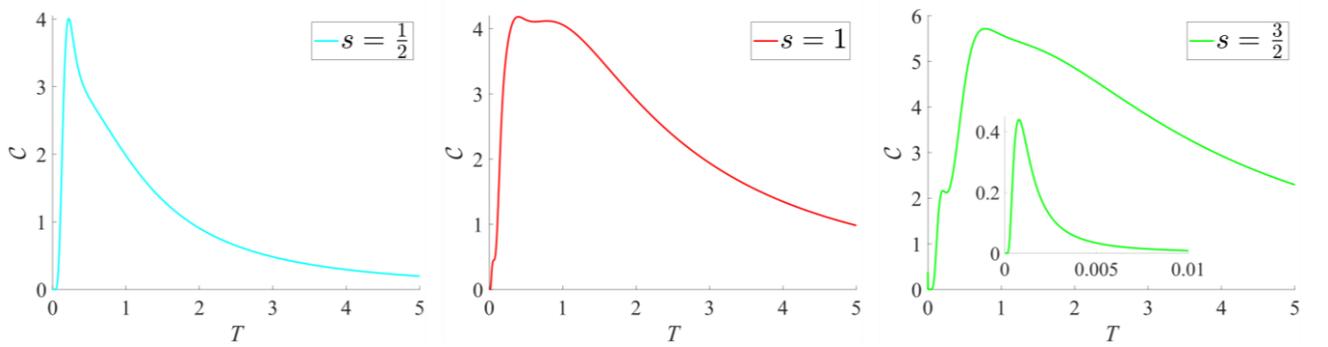

Figure 7: Zero-field magnetic heat-capacity for the icosahedron with $s = \frac{1}{2}$, $s = 1$ and $s = \frac{3}{2}$. A low-temperature peak is observed at $T \approx 0.008$ for $s = \frac{3}{2}$ (inset), reflecting the near-degeneracy of the ground-state.

**Cuboctahedron.** A comparison between diagonalization in $(\Gamma, M)$-and $(\Gamma, S)$-spaces for the $s = 1$ cuboctahedron, as shown in Figure 8, again demonstrates good agreement.

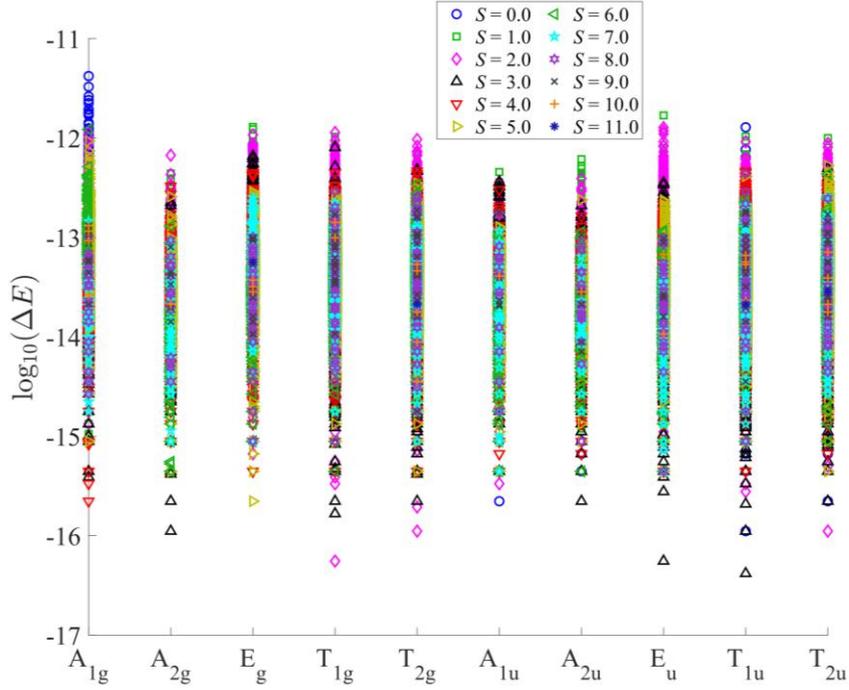

Figure 8: Comparison of energy differences from diagonalization based on adaptation to $(\Gamma, M)$ or $(\Gamma, S)$, as a function of the irreps $\Gamma$ of the $O_h$ point-group, for an $s=1$ cuboctahedron.

The $s = \frac{3}{2}$ system was solved by Schnalle and Schnack in the compatible $D_2$-subgroup and only the low-energy part of the spectrum was fully resolved with respect to $O_h$ [6,15]. Here, we have for the first time obtained a full symmetry classification of the entire spectrum (the subspace-dimensions are given in Table 4 in the Appendix), see Figure 9.

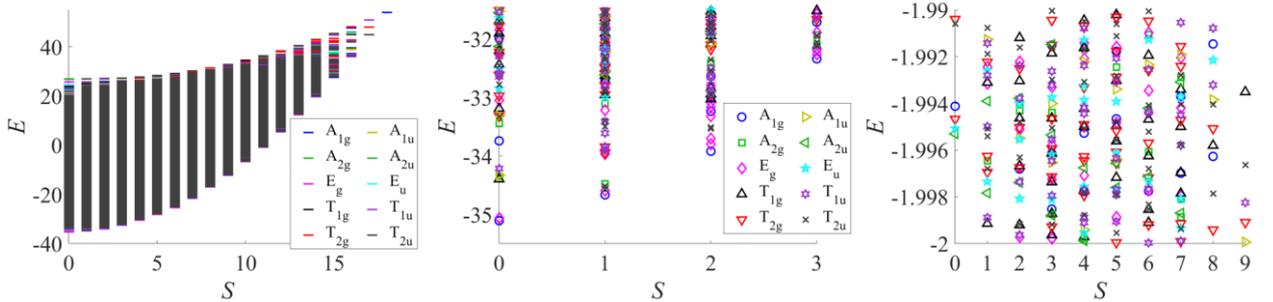

Figure 9: Full spectrum of the $s = \frac{3}{2}$ cuboctahedron (left), low-energy section (middle), and an arbitrary section from the intermediate energy-range (right).

We plot the magnetic heat-capacities in Figure 10. A comparison with Ref. [6] reveals that our calculated curves coincide with theirs if the temperature-axis is scaled by a factor of two. This difference could arise from a distinct choice of conventions in defining the exchange constant $J$ in the Heisenberg Hamiltonian. However, our convention for energies matches those in Ref. [6], as apparent from Figs. 5 and 6 in that work, implying an inconsistency in the treatment of $J$-conventions.

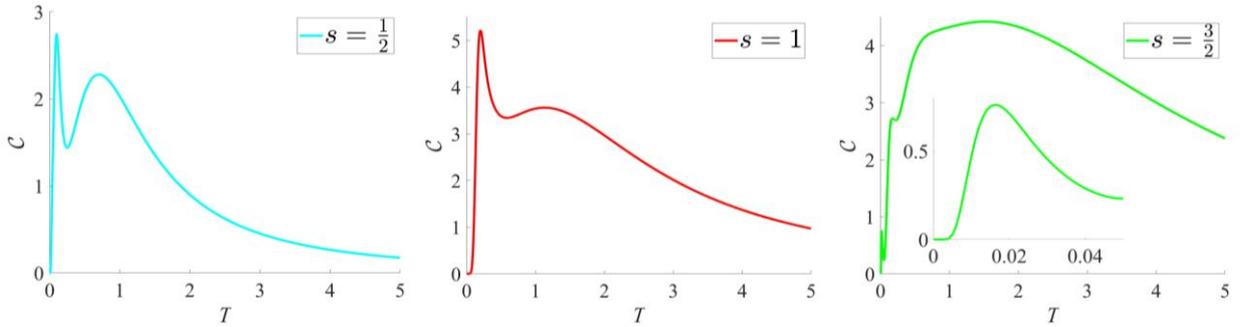

Figure 10: Zero-field magnetic heat-capacity for the cuboctahedron with $s=\tfrac{1}{2}$, $s=1$ and $s=\tfrac{3}{2}$. A low-temperature peak at $T \approx 0.016$ for $s=\tfrac{3}{2}$ (inset) is due to the near-degeneracy of the ground-state (cf. Figure 9).

**Truncated Tetrahedron.** Finally, we consider the truncated tetrahedron, which, to the best of our knowledge, has never been fully diagonalized for $s=\tfrac{3}{2}$. This may be attributed to the fact that the relevant $T_d$ point-group (which is isomorphic to the octahedral group $O$) with only 5 irreps comprising a total of 10 components is smaller than $O_h$ or $I_h$ with 10 irreps each and 20 and 30 components, respectively. Consequently, the matrices involved are significantly larger but still manageable if the full symmetry is exploited (see Table 5 in the Appendix). The spectrum of the $s=\tfrac{3}{2}$ truncated tetrahedron is presented in Figure 11, and the heat-capacities for the $s=\tfrac{1}{2}$, $s=1$ and $s=\tfrac{3}{2}$ systems are shown in Figure 12.

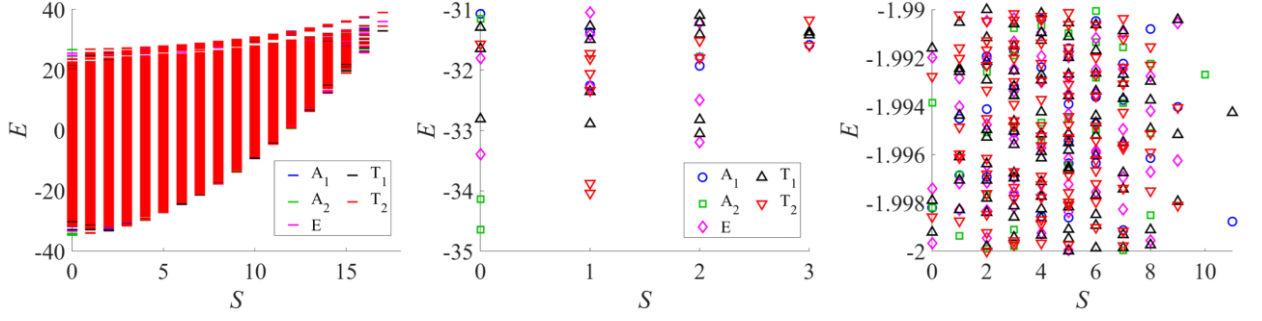

Figure 11: Full spectrum of the $s=\frac{3}{2}$ truncated tetrahedron (left), low-energy section (middle), and an arbitrary section from the intermediate energy-range (right).

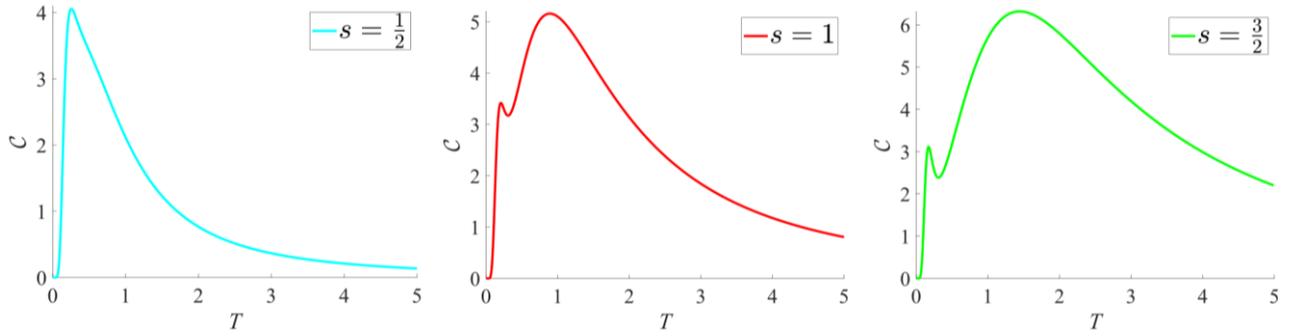

Figure 12: Zero-field magnetic heat-capacity for the truncated tetrahedron with $s=\frac{1}{2}$, $s=1$ and $s=\frac{3}{2}$.

## 4. Summary and Conclusion

We have presented a simple and efficient approach for the exact diagonalization (ED) of isotropic spin-clusters by combining spin and point-group (PG) symmetries. This method applies projection-operators to uncoupled basis-states, thereby eliminating the need for the complex recoupling-transformations typically required in schemes based on genealogical coupling (GC). The accuracy and versatility of the present projection-approach, and its ability to handle systems that are computationally inaccessible with other techniques, was demonstrated on spin-rings and polyhedra. Notably, we could fully determine and classify the spectra of the $s=\frac{3}{2}$ icosahedron and cuboctahedron, which had previously been only partially resolved due to computational limitations imposed by recoupling-transformations, which are avoided in the present projection-approach. Moreover, we diagonalized the $s=\frac{3}{2}$ truncated tetrahedron for the first time, a task that would pose a significant challenge without a full exploitation of symmetry. Due to its numerical efficiency, conceptual simplicity and straightforward implementation, the projection-method potentially represents a standard for tackling large Heisenberg spin-clusters via exact diagonalization.

**Acknowledgements.** We thank TU Berlin for providing computational resources.

# Appendix

Here, we compile the dimensions symmetry-subspaces for some systems discussed in this work, specifically the icosahedron, cuboctahedron and truncated tetrahedron.

Table 1: Dimensions of combined total-spin and $I_h$ point-group subspaces for the $s = 1$ icosahedron.

| S | $A_g$ | $A_u$ | $T_{1g}$ | $T_{1u}$ | $T_{2g}$ | $T_{2u}$ | $F_g$ | $F_u$ | $H_g$ | $H_u$ |
|---|---|---|---|---|---|---|---|---|---|---|
| 0 | 63 | 50 | 87 | 89 | 87 | 89 | 148 | 138 | 200 | 180 |
| 1 | 84 | 74 | 289 | 308 | 289 | 308 | 372 | 380 | 456 | 454 |
| 2 | 153 | 124 | 350 | 370 | 350 | 370 | 502 | 494 | 663 | 626 |
| 3 | 127 | 104 | 370 | 399 | 370 | 399 | 498 | 504 | 617 | 600 |
| 4 | 127 | 96 | 278 | 303 | 278 | 303 | 406 | 400 | 533 | 496 |
| 5 | 71 | 50 | 198 | 223 | 198 | 223 | 268 | 272 | 345 | 326 |
| 6 | 58 | 36 | 101 | 119 | 101 | 119 | 158 | 154 | 210 | 186 |
| 7 | 22 | 10 | 51 | 65 | 51 | 65 | 73 | 74 | 95 | 84 |
| 8 | 14 | 4 | 15 | 23 | 15 | 23 | 29 | 28 | 45 | 34 |
| 9 | 4 | 0 | 5 | 10 | 5 | 10 | 10 | 10 | 12 | 8 |
| 10 | 3 | 0 | 0 | 2 | 0 | 2 | 2 | 2 | 5 | 2 |
| 11 | 0 | 0 | 0 | 1 | 0 | 1 | 0 | 1 | 0 | 0 |
| 12 | 1 | 0 | 0 | 0 | 0 | 0 | 0 | 0 | 0 | 0 |

Table 2: Dimensions of combined $S_z$ and $I_h$ point-group subspaces for the $s = \frac{3}{2}$ icosahedron.

| M | $A_g$ | $A_u$ | $T_{1g}$ | $T_{1u}$ | $T_{2g}$ | $T_{2u}$ | $F_g$ | $F_u$ | $H_g$ | $H_u$ |
|---|---|---|---|---|---|---|---|---|---|---|
| 0 | 14920 | 13640 | 41900 | 43140 | 41900 | 43140 | 56812 | 56772 | 71708 | 70392 |
| 1 | 14382 | 13136 | 40648 | 41894 | 40648 | 41894 | 55024 | 55024 | 69406 | 68160 |
| 2 | 13179 | 11988 | 36895 | 38049 | 36895 | 38049 | 50032 | 50032 | 63247 | 62020 |
| 3 | 11226 | 10146 | 31493 | 32573 | 31493 | 32573 | 42714 | 42714 | 53920 | 52840 |
| 4 | 9076 | 8116 | 25091 | 26020 | 25091 | 26020 | 34162 | 34132 | 43238 | 42248 |
| 5 | 6792 | 5984 | 18754 | 19562 | 18754 | 19562 | 25540 | 25540 | 32332 | 31524 |
| 6 | 4840 | 4174 | 13012 | 13654 | 13012 | 13654 | 17847 | 17824 | 22670 | 21984 |
| 7 | 3156 | 2640 | 8432 | 8948 | 8432 | 8948 | 11584 | 11584 | 14740 | 14224 |
| 8 | 1968 | 1578 | 5020 | 5395 | 5020 | 5395 | 6984 | 6970 | 8952 | 8548 |
| 9 | 1112 | 836 | 2771 | 3047 | 2771 | 3047 | 3880 | 3880 | 4982 | 4706 |
| 10 | 605 | 416 | 1377 | 1558 | 1377 | 1558 | 1978 | 1970 | 2583 | 2386 |
| 11 | 290 | 170 | 627 | 747 | 627 | 747 | 914 | 914 | 1204 | 1084 |
| 12 | 142 | 68 | 245 | 314 | 245 | 314 | 384 | 380 | 520 | 444 |
| 13 | 56 | 16 | 86 | 126 | 86 | 126 | 140 | 140 | 196 | 156 |
| 14 | 25 | 4 | 22 | 41 | 22 | 41 | 45 | 44 | 70 | 48 |
| 15 | 9 | 0 | 5 | 14 | 5 | 14 | 12 | 12 | 19 | 10 |
| 16 | 4 | 0 | 0 | 3 | 0 | 3 | 2 | 2 | 6 | 2 |
| 17 | 1 | 0 | 0 | 1 | 0 | 1 | 0 | 0 | 1 | 0 |
| 18 | 1 | 0 | 0 | 0 | 0 | 0 | 0 | 0 | 0 | 0 |

Table 3: Dimensions of combined total-spin and $I_h$ point-group subspaces for the $s = \frac{3}{2}$ icosahedron.

| S | A$_g$ | A$_u$ | T$_{1g}$ | T$_{1u}$ | T$_{2g}$ | T$_{2u}$ | F$_g$ | F$_u$ | H$_g$ | H$_u$ |
|---|---|---|---|---|---|---|---|---|---|---|
| 0 | 538 | 504 | 1252 | 1246 | 1252 | 1246 | 1788 | 1748 | 2302 | 2232 |
| 1 | 1203 | 1148 | 3753 | 3845 | 3753 | 3845 | 4956 | 4992 | 6159 | 6140 |
| 2 | 1953 | 1842 | 5402 | 5476 | 5402 | 5476 | 7354 | 7318 | 9327 | 9180 |
| 3 | 2150 | 2030 | 6402 | 6553 | 6402 | 6553 | 8552 | 8582 | 10682 | 10592 |
| 4 | 2284 | 2132 | 6337 | 6458 | 6337 | 6458 | 8622 | 8592 | 10906 | 10724 |
| 5 | 1952 | 1810 | 5742 | 5908 | 5742 | 5908 | 7693 | 7716 | 9662 | 9540 |
| 6 | 1684 | 1534 | 4580 | 4706 | 4580 | 4706 | 6263 | 6240 | 7930 | 7760 |
| 7 | 1188 | 1062 | 3412 | 3553 | 3412 | 3553 | 4600 | 4614 | 5788 | 5676 |
| 8 | 856 | 742 | 2249 | 2348 | 2249 | 2348 | 3104 | 3090 | 3970 | 3842 |
| 9 | 507 | 420 | 1394 | 1489 | 1394 | 1489 | 1902 | 1910 | 2399 | 2320 |
| 10 | 315 | 246 | 750 | 811 | 750 | 811 | 1064 | 1056 | 1379 | 1302 |
| 11 | 148 | 102 | 382 | 433 | 382 | 433 | 530 | 534 | 684 | 640 |
| 12 | 86 | 52 | 159 | 188 | 159 | 188 | 244 | 240 | 324 | 288 |
| 13 | 31 | 12 | 64 | 85 | 64 | 85 | 95 | 96 | 126 | 108 |
| 14 | 16 | 4 | 17 | 27 | 17 | 27 | 33 | 32 | 51 | 38 |
| 15 | 5 | 0 | 5 | 11 | 5 | 11 | 10 | 10 | 13 | 8 |
| 16 | 3 | 0 | 0 | 2 | 0 | 2 | 2 | 2 | 5 | 2 |
| 17 | 0 | 0 | 0 | 1 | 0 | 1 | 0 | 0 | 1 | 0 |
| 18 | 1 | 0 | 0 | 0 | 0 | 0 | 0 | 0 | 0 | 0 |

Table 4: Dimensions of combined total-spin and $O_h$ point-group subspaces for the $s = \frac{3}{2}$ cuboctahedron.

| S | A$_{1g}$ | A$_{2g}$ | E$_g$ | T$_{1g}$ | T$_{2g}$ | A$_{1u}$ | A$_{2u}$ | E$_u$ | T$_{1u}$ | T$_{2u}$ |
|---|---|---|---|---|---|---|---|---|---|---|
| 0 | 1163 | 1163 | 2302 | 3297 | 3297 | 1126 | 1126 | 2232 | 3236 | 3236 |
| 1 | 3095 | 3064 | 6159 | 9289 | 9332 | 3070 | 3070 | 6140 | 9411 | 9411 |
| 2 | 4675 | 4632 | 9327 | 13727 | 13758 | 4580 | 4580 | 9180 | 13725 | 13725 |
| 3 | 5382 | 5320 | 10682 | 15988 | 16050 | 5306 | 5306 | 10592 | 16140 | 16140 |
| 4 | 5484 | 5422 | 10906 | 16070 | 16132 | 5362 | 5362 | 10724 | 16116 | 16116 |
| 5 | 4854 | 4791 | 9662 | 14383 | 14456 | 4763 | 4763 | 9540 | 14536 | 14536 |
| 6 | 4010 | 3937 | 7930 | 11645 | 11708 | 3887 | 3887 | 7760 | 11706 | 11706 |
| 7 | 2923 | 2865 | 5788 | 8577 | 8635 | 2838 | 2838 | 5676 | 8698 | 8698 |
| 8 | 2009 | 1951 | 3970 | 5757 | 5815 | 1916 | 1916 | 3842 | 5814 | 5814 |
| 9 | 1223 | 1186 | 2399 | 3523 | 3566 | 1165 | 1165 | 2320 | 3604 | 3604 |
| 10 | 711 | 668 | 1379 | 1953 | 1990 | 651 | 651 | 1302 | 1990 | 1990 |
| 11 | 350 | 328 | 684 | 978 | 1000 | 318 | 318 | 640 | 1020 | 1020 |
| 12 | 176 | 154 | 324 | 432 | 454 | 146 | 146 | 288 | 452 | 452 |
| 13 | 67 | 59 | 126 | 169 | 180 | 54 | 54 | 108 | 187 | 187 |
| 14 | 30 | 19 | 51 | 55 | 63 | 18 | 18 | 38 | 62 | 62 |
| 15 | 9 | 6 | 13 | 15 | 18 | 5 | 5 | 13 | 20 | 20 |
| 16 | 4 | 1 | 5 | 2 | 5 | 1 | 1 | 5 | 4 | 4 |
| 17 | 0 | 0 | 1 | 0 | 1 | 0 | 0 | 1 | 1 | 1 |
| 18 | 1 | 0 | 0 | 0 | 0 | 0 | 0 | 0 | 0 | 0 |

Table 5: Dimensions of combined total-spin and $T_d$ point-group subspaces for the $s=\frac{3}{2}$ truncated tetrahedron.

| S | A$_1$ | A$_2$ | E | T$_1$ | T$_2$ |
|---|---|---|---|---|---|
| 0 | 2289 | 2289 | 4534 | 6533 | 6533 |
| 1 | 6165 | 6134 | 12299 | 18700 | 18743 |
| 2 | 9255 | 9212 | 18507 | 27452 | 27483 |
| 3 | 10688 | 10626 | 21274 | 32128 | 32190 |
| 4 | 10846 | 10784 | 21630 | 32186 | 32248 |
| 5 | 9617 | 9554 | 19202 | 28919 | 28992 |
| 6 | 7897 | 7824 | 15690 | 23351 | 23414 |
| 7 | 5761 | 5703 | 11464 | 17275 | 17333 |
| 8 | 3925 | 3867 | 7812 | 11571 | 11629 |
| 9 | 2388 | 2351 | 4719 | 7127 | 7170 |
| 10 | 1362 | 1319 | 2681 | 3943 | 3980 |
| 11 | 668 | 646 | 1324 | 1998 | 2020 |
| 12 | 322 | 300 | 612 | 884 | 906 |
| 13 | 121 | 113 | 234 | 356 | 367 |
| 14 | 48 | 37 | 89 | 117 | 125 |
| 15 | 14 | 11 | 21 | 35 | 38 |
| 16 | 5 | 2 | 7 | 6 | 9 |
| 17 | 0 | 0 | 1 | 1 | 2 |
| 18 | 1 | 0 | 0 | 0 | 0 |